\documentclass[prl,showpacs, twocolumn]{revtex4}

\usepackage{graphicx}

\begin{document}
\title{Observation of Topological and Parity-dependent Phase of $m=0$ Spin States}

\author{Koji Usami$^{1,2}$} \email{usami.k.ab@m.titech.ac.jp}
\author{Mikio Kozuma$^{2,3}$}  
\affiliation{$^1$PRESTO, Japan Science and Technology Agency, 3-5, Sanbancho, Chiyodaku, Tokyo 332-0012, Japan \\
$^2$Department of Physics, Tokyo Institute of Technology, 2-12-1 O-okayama, Meguro-ku, Tokyo 152-8550, Japan \\
$^3$CREST, Japan Science and Technology Agency, 3-5, Sanbancho, Chiyodaku, Tokyo 332-0012, Japan}
\date{\today }

\begin{abstract}
A Ramsey interrogation scheme was used to measure the phase shift of laser-cooled $^{87}$Rb clock-transition pseudospins arising as a result of a reversal of a bias magnetic field, i.e., $\textbf{B} \to -\textbf{B}$, during the interrogation. While no phase shift occurred when the reversal was sudden, the Ramsey fringes were shifted by a factor of $\pi$ when the reversal was adiabatic. We thus verified the prediction that the spin states $|F,m=0 \rangle$ acquire a purely topological and parity-dependent phase factor of $(-1)^{F}$ as a result of $\textbf{B} \to -\textbf{B}$.
\end{abstract}


\maketitle

One of the most astonishing prediction of quantum mechanics is the $\pi$-phase shift of spin-1/2 particles by $2 \pi$ rotations~\cite{Sakurai}. The non-trivial $\pi$-phase shift was claimed to be measurable~\cite{AS1967pr,Bernstein1967}, and later proved by the beautiful neutron interference experiments~\cite{RZBWBB1975pla,WCOE1975}. The phase shift can be interpreted by the double-valuedness of the spinor wave function under three-dimensional rotations~\cite{Zare}.

In this Letter, we report on an experiment on the non-trivial $\pi$-phase shift of a seemingly trivial integer spin. To see its curious behavior, let us consider the following situation: an atom with integer total spin, $F$ (here, we assume the orbital angular momentum $l=0$ for brevity), and its z-component, $m$, goes through a reversal of a bias magnetic field from, say, $\textbf{B}(0)=(0,0,B_{z})$ at time $t=0$ to $\textbf{B}(T)=-\textbf{B}(0)=(0,0,-B_{z})$ at $t=T$. When the bias magnetic field is kept finite in the course of the reversal (i.e.,$|\textbf{B}(t)| \neq 0$) and changes sufficiently slowly as to avoid energy level crossing and satisfy the adiabatic condition, each spin eigenstate remains as an instantaneous one~\cite{Kato,Messiah}. Here, the peculiar thing about the $m=0$ eigenstate is that the final state at $t=T$ is exactly the same state as the initial eigenstate $|F,m=0 \rangle$ up to a certain phase factor. Then, the question is ``what phase factor arises?" Quantum mechanics tells us that the phase is $(-1)^{F}$ as is derived as follows: Without loss of generality, the adiabatic reversal $(0,0,B_{z}) \to (0,0,-B_{z})$ with $|\textbf{B}(t)| \neq 0$ is equivalent to a y-axis [or x-axis] $\pi$ rotation of the $|F,m=0 \rangle$, i.e., $Y_{F0}(\theta,\phi) \to Y_{F0}(\pi-\theta,\pi-\phi)$ [or $Y_{F0}(\theta,\phi) \to Y_{F0}(\pi-\theta,2\pi-\phi)$], where $Y_{F0}(\theta,\phi)=\langle \theta,\phi|F,m=0 \rangle$ are the spherical harmonics and $(\theta,\phi) \to (\pi-\theta,\pi-\phi)$ [or $(\theta,\phi) \to(\pi-\theta,2\pi-\phi)$] corresponds to a y-axis [or x-axis] $\pi$ rotation. Since the spherical harmonics for $m=0$, i.e., $Y_{F0}(\theta,\phi)$, do not depend on $\phi$, then the y-axis [or x-axis] $\pi$ rotation is identical to the space inversion $(\theta,\phi) \to (\pi-\theta,\pi+\phi)$. Thus, we have the aforesaid phase factor:
\begin{equation}
Y_{F0}(\theta,\phi) \to Y_{F0}(\pi-\theta,\pi+\phi)=(-1)^{F}Y_{F0}(\theta,\phi), \label{e:phase}
\end{equation} 
from the parity of $Y_{F0}(\theta,\phi)$~\cite{Sakurai,Zare}.

The phase factor of $(-1)^{F}$ in Eq.~(\ref{e:phase}) is interesting in the following three points: First, the phase is neither a dynamical phase nor a conventional geometric phase~\cite{Berry1984prsla}. The dynamical phase is $0$ because the Zeeman energy, $E= g_{F}\mu_{B} \textbf{F} \cdot \textbf{B}(t)$, remains $0$ for the $m=0$ eigenstate throughout the course of the bias-magnetic-field reversal, where $g_{F}$ is the Land\'{e} factor and $\mu_{B}$ is Bohr magneton. The conventional geometric phase is also $0$ because the ``magnetic field" induced by the parameter space $\textbf{B}(t)$~\cite{Berry1984prsla} (in other words, Berry's curvature~\cite{Simon1983,Nakahara}) is $0$ for the $m=0$ eigenstate. Robbins and Berry, in their imaginative paper~\cite{RB1994jpa}, gave a topological explanation for the phase factor of $(-1)^{F}$. The underlying topology is elucidated by considering the parameter space $\textbf{B}(t)$ for the $m=0$ eigenstate, in which each point determines the $m=0$ eigenstate up to a phase, as a real projective plane~\cite{Nakahara}. Here, $|\textbf{B}| \neq 0$ and the point $\textbf{B}$ can be identified with the antipodal point $-\textbf{B}$. The non-trivial closed cycle on the real projective plane results in the phase factor of $(-1)^{F}$~\cite{RB1994jpa}. The phase, $(-1)^{F}$, is the key to a novel non-relativistic account of the spin-statistics connection~\cite{BR1997prsla}. Note that the phase has just two values, $\{0,\pi\}$, and should be robust against perturbations of the closed cycle paths due to its topological character. 

Second, despite the above distinction between the conventional geometric phase~\cite{Berry1984prsla} and the topological phase there is a similarity: the phase is associated with the level crossing at $\textbf{B}=(0,0,0)$, which is in fact a hallmark of the former~\cite{Berry1984prsla,Fujikawa2005mpla}. Thus, to obtain the phase factor of $(-1)^{F}$, the bias-magnetic-field reversal must be adiabatic, otherwise the parameter space $\textbf{B}(t)$ becomes topologically trivial~\cite{Fujikawa2005mpla}.

Third, the phase, $(-1)^{F}$, is parity-dependent. This is a remarkable consequence of quantum mechanics. As mentioned earlier, the reversal of a bias magnetic field is identical to the space inversion $(\theta,\phi) \to (\pi-\theta,\pi+\phi)$ for the $\phi$-independent $m=0$ eigenstate, $Y_{F0}(\theta,\phi)$. We can say that this is another aspect of the real projective plane. Thus the $m=0$ eigenstate can go to the mirror world just by rotating $\pi$ around the y axis (or x axis). Consequently, whether the phase is non-trivial $\pi$ or trivial $0$ depends on whether $Y_{F0}(\theta,\phi)$ has odd or even parity.

The principle of the topological phase measurement is based on a Ramsey interrogation scheme~\cite{Foot} for a laser-cooled $^{87}$Rb clock-transition pseudospin that consists of $|5^{2}S_{1/2}\ F=1,m=0 \rangle$ and $|5^{2}S_{1/2}\ F=2,m=0 \rangle$. The first microwave $\pi/2$ pulse tuned to the $^{87}$Rb hyperfine splitting rotates the clock-transition pseudospin, say, from the initial state $|F=2,m=0 \rangle$ to the superposition state $1/\sqrt{2}(|F=2,m=0 \rangle+|F=1,m=0 \rangle)$ in the bias magnetic field $\textbf{B}$ along the z axis. When the detuning, $\Delta$, between the hyperfine splitting $\nu_{hf}$ and the microwave frequency $\nu$ is zero, the clock-transition pseudospin remains in the above superposition state in a frame rotating at $\nu$ as long as the coherence of the pseudospin is preserved. The second microwave $\pi/2$ pulse rotates the pseudospin to $|F=1,m=0 \rangle$, thus the pseudospin is completely flipped. By changing the detuning, $\Delta$, and probing the final population of, say, $|F=2,m=0 \rangle$, we can obtain the celebrated Ramsey fringes~\cite{Foot}. This is in a sense a measurement of the dynamical phase due to the hyperfine splitting with the $\Delta$-detuned microwave field. If, however, the bias magnetic field is adiabatically reversed during the interrogation time, i.e., the time between two resonant $\pi/2$ pulses, the states $|F=1,m=0 \rangle$ and $|F=2,m=0 \rangle$ acquire the topological phases $-1$ and $1$, respectively. The resultant superposition state is then $1/\sqrt{2}(|F=2,m=0 \rangle-|F=1,m=0 \rangle)$. In this case we have the state $|F=2,m=0 \rangle$ after the second microwave $\pi/2$ pulse. Accordingly, the topological and parity-dependent phase $(-1)^{F}$ manifests itself as the $\pi$-phase shifted Ramsey fringes.

\begin{figure}
\includegraphics[width=0.9\linewidth]{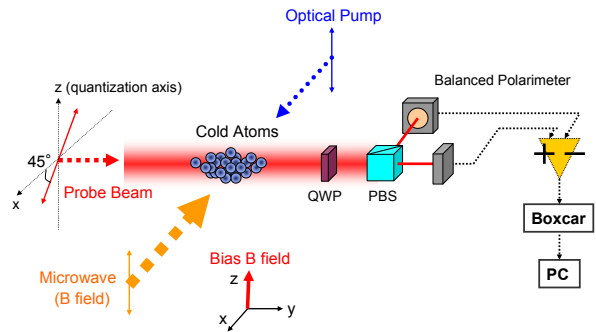}
\caption{Schematic of the apparatus. Cold atoms of $^{87}$Rb in the $|F=2,m=0 \rangle$ hyperfine state were prepared with a magneto-optical trap (MOT) and optical pumping. A probe beam with linear polarization tilted by 45$^{\circ}$ relative to the x axis propagates through the atomic cloud along the y axis. Linear birefringent phase shifts were detected using a balanced polarimeter with a quarter-wave plate (QWP) and a polarization beam splitter (PBS) and stored in a PC with a boxcar integrator and an A/D converter. A microwave field at ground-state hyperfine splitting was used for manipulating the clock-transition pseudospins.}
\label{fig:schematic}
\end{figure}

A schematic of the apparatus used in our measurements is shown in Fig.~\ref{fig:schematic}. One cycle of measurement consisted of a cooling period (35~ms) and a probing period (5~ms). We began by preparing cold atoms of $^{87}$Rb with a magneto-optical trap (MOT)~\cite{Foot}. The quadrupole magnetic field for the MOT was on only for the first 25~ms in the cooling period to reduce the eddy-current-induced magnetic field in the probing period~\cite{IKYMKK2006a}. The cooling laser was switched off at the end of the cooling period. The probing period began by optically pumping the atoms into the $|F=2,m=0 \rangle$ hyperfine state by using a 500-$\mu$W $\pi$-polarized laser tuned to the $5^{2}S_{1/2}\ F=2 \to 5^{2}P_{3/2}\ F'=2$ transition during 150~$\mu$s with the repumping laser tuned to the $5^{2}S_{1/2}\ F=1 \to 5^{2}P_{3/2}\ F'=2$ transition and a bias magnetic field ($\sim$~200~mG) along the z axis as shown in Fig.~\ref{fig:schematic}. All the measurements were performed within the last 4~ms in the probing period. Note that during the probing period the stray magnetic fields around the atomic cloud were actively canceled down to less than 1~mG with three magneto-impedance (MI) sensors (Aichi Micro Intelligent)~\cite{PM1994apl} and three pairs of compensating coils for the respective directions~\cite{RSG2001a}.

\begin{figure*}
\begin{center}
\includegraphics[width=0.85\linewidth]{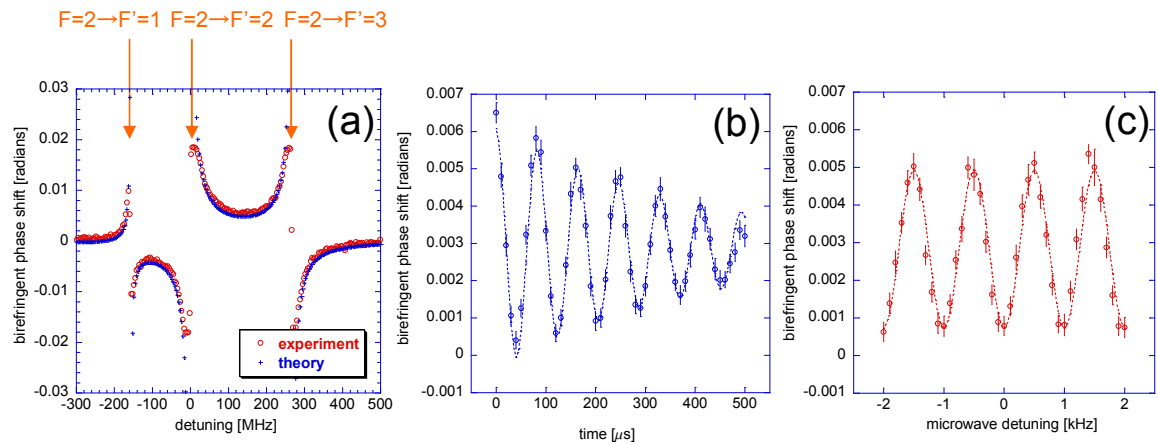}
\caption{(a) Measured linear birefringent phase shift as a function of the detuning of the probe laser from $5^{2}S_{1/2}\ F=2 \to 5^{2}P_{3/2}\ F'=2$ transition. (b) Rabi oscillations of the clock-transition pseudospins. (c) Ramsey fringes. Each point and error bar in (b) and (c) corresponds to the average and the $\pm$~1 standard deviation of 100 birefringence measurements, respectively.}
\label{fig:result1}
\end{center}
\end{figure*}

Figure~\ref{fig:result1}~(a) shows the linear birefringent phase shift as a function of the detuning of the probe laser from $5^{2}S_{1/2}\ F=2 \to 5^{2}P_{3/2}\ F'=2$ transition. Here, a 1-$\mu$s probe pulse with a diameter of $\sim$~3~mm and a power of 30 $\mu$W propagated through the atomic cloud along the y axis as shown in Fig.~\ref{fig:schematic} and underwent linear birefringent phase shift between x and z polarization components, that is proportional to the quadruple moment of the atomic angular momentum~\cite{GSM2006a,CSSJ2006}. The linear birefringent phase shifts were detected using a shot-noise-limited time-domain balanced polarimeter~\cite{HAHLLMS2001op} with a quarter-wave plate (QWP) and a polarization beam splitter (PBS) and stored in a PC with a boxcar integrator (SRS SR250) and an A/D converter. The optical density of the atomic cloud was roughly estimated to be 2.0 from the fit of the data to the theoretical calculation based on ref.~\cite{GSM2006a,CSSJ2006}.

Figure~\ref{fig:result1}~(b) shows the Rabi oscillations of the clock-transition pseudospins, which were obtained by the birefringent phase-shift measurement under irradiation of the resonant microwave field at the $^{87}$Rb ground-state hyperfine splitting~\cite{BSSMCS1999epl}. To generate the microwave field with the required precision (for our experiment at least 10~Hz), a rubidium frequency standard (SRS FS725) was used. The 10-MHz reference signal was fed into a 5.8-GHz phase locked signal source (Herley-CTI) and a frequency-variable (up to 1.1~GHz) signal generator (R\&S SML01). After mixing the above two signals, filtering out the spurious frequency components, and amplifying the desired frequency components, we converted the signal in the coaxial cable into one in the waveguide and then one in free space with a home-made microwave horn. The resulting microwave field ($\sim$ 32~dBm) was then focused on the atomic cloud with a 20~$\times$~15~cm concave gold-coating microwave mirror. Before the amplifier, there was a PIN diode switch (UMCC) for microwave-pulse clipping. Each point in Fig.~\ref{fig:result1}~(b) corresponds to the average of 100 birefringence measurements. Here, the probe laser was detuned to 100-MHz blue from the $5^{2}S_{1/2}\ F=2 \to 5^{2}P_{3/2}\ F'=2$ transition to diminish any absorptions yet to obtain appreciable birefringence exclusively due to the atoms in the state, $|F=2,m=0 \rangle$. Thus, we could measure the population of $|F=2,m=0 \rangle$ without destroying the z component of the angular momentum of the collective pseudospin~\cite{CSSJ2006} and thus with a excellent signal-to-noise ratio. The observed Rabi frequency was about 12.2~kHz and the decay time was about 0.4~ms, which may be primarily due to the inhomogeneity of the microwave power over the atomic cloud. Without irradiation of the microwave field, the decay time was about 5.5~ms and may be interpreted as resulting from atomic diffusion. 

\begin{figure}
\includegraphics[width=\linewidth]{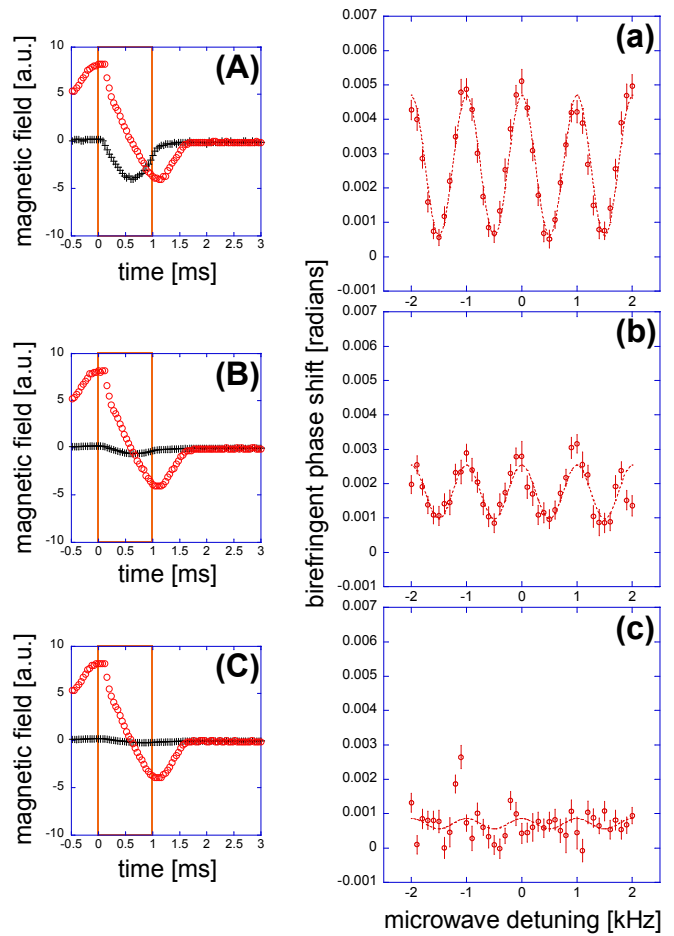}
\caption{(A), (B), and (C): Measured traces of the bias magnetic fields along the z axis (circle) and x axis (cross) for the adiabatic reversals. The Ramsey interrogation periods are indicated by boxes. (a), (b), and (c): Resultant Ramsey fringes.}
\label{fig:result2}
\end{figure}

Figure~\ref{fig:result1}~(c) shows the Ramsey fringes, which were obtained by the birefringent phase-shift measurement after the second microwave $\pi/2$-pulse irradiation. Each point corresponds to the average of 100 birefringence measurements with a given microwave detuning, $\Delta$. Here, the interrogation time, i.e., the time between two $\pi/2$ pulses, was 1~ms, which resulted in the fringe width of 1~kHz. The bias magnetic field of $\sim$~200~mG along the z axis and the active magnetic-field cancellation for the other axes prevented the atoms in $m=0$ states from changing to the other Zeeman sublevels during the interrogation.   

\begin{figure}
\includegraphics[width=\linewidth]{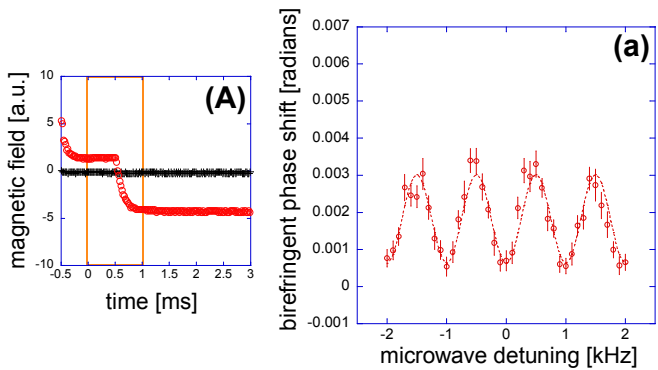}
\caption{(A) Measured traces of the bias magnetic fields along z axis (circle) and x axis (cross) for the sudden reversal. The Ramsey interrogation period is indicated by a box. (a) Resultant Ramsey fringes.}
\label{fig:result3}
\end{figure}

Figures~\ref{fig:result2}~(a), (b), (c), and \ref{fig:result3}~(a) show the Ramsey fringes as a result of reversals of the bias magnetic fields during the interrogation. The ways in which the bias magnetic fields were reversed are presented in Figs.~\ref{fig:result2}~(A), (B), (C), and \ref{fig:result3}~(A), which show the real traces of the magnetic fields along the z axis (circle) and x axis (cross) probed by the MI sensors (one unit on the vertical scale is roughly equivalent to 40~mG). Here, the interrogation periods are indicated by boxes. Whether the bias-magnetic-field reversal is adiabatic or sudden can be crudely determined by whether $\Delta \tau \gg \hbar /\Delta E$ or $\Delta \tau \ll \hbar /\Delta E$~\cite{Messiah,Fujikawa2005mpla}, where $\Delta \tau$ is the typical time scale of the bias-magnetic-field reversal and $\Delta E$ is the minimum energy gap between $m=0$ Zeeman sublevel and the nearest one during the reversal. Thus, the reversals shown in Figs.~\ref{fig:result2}~(A), (B), and (C) were supposed to be adiabatic since $\Delta \tau \sim$~2~ms and $\Delta E/h \sim$~140~kHz, 14~kHz, and 2.8~kHz, respectively; whereas the reversal shown in Fig.~\ref{fig:result3}~(A) was supposed to be sudden since $\Delta \tau \sim$~2~$\mu$s and $\Delta E/h \sim$~700~Hz (due to the residual magnetic fields about 1~mG). The $\pi$-phase shifted Ramsey fringes in Fig.~\ref{fig:result2}~(a) evidently show that the clock-transition pseudospins acquired the topological and parity-dependent phase $(-1)^{F}$ as a result of the adiabatic bias-magnetic-field reversal. On the other hand, the Ramsey fringes in Fig.~\ref{fig:result3}~(a) were, as expected, not shifted from the original ones in Fig.~\ref{fig:result1}~(c) since the bias-magnetic-field reversal was sudden. Note that to obtain the Ramsey fringes in Fig.~\ref{fig:result3}~(a), we adjusted the magnetic fields of the x and y axes to be zero with great care, otherwise the residual magnetic fields during the interrogation would have caused the mixing of the Zeeman sublevels and thus resulted in the disappearance of the Ramsey fringes. By the same token, the fringe visibilities in Figs.~\ref{fig:result2}~(b) and (c) worsened as the minimum values of the bias magnetic fields during the interrogation became small. Nevertheless, we can observe the $\pi$-phase shifted Ramsey fringes plainly in Fig.~\ref{fig:result2}~(b) and barely in Fig.~\ref{fig:result2}~(c) showing the robustness of the topological phase \textit{per se} against perturbations.

In conclusion, we have observed the topological and parity-dependent phase shift of laser-cooled $^{87}$Rb clock-transition pseudospins as a result of a reversal of a bias magnetic field with the Ramsey interrogation scheme. The distinctive features of the phase, i.e., its robustness against perturbations and the relation to adiabaticity, have been confirmed. The unique character of the phase is its parity dependence, which may provide a new scheme for testing parity non-conservation in atoms~\cite{WBCMRTW1997s}.

\begin{acknowledgments}
We thank Keiichirou~Akiba, Kazuhito~Honda, Akio~Hosoya, Hideto~Kanamori, Kousuke~Kashiwagi, Yuki~Kawaguchi, Keiji Matsumoto, Makoto~Takeuchi, Hiroaki~Terashima, and Masahito~Ueda for fruitful discussions. We also acknowledge the generous loan of a microwave mirror by Hideto~Kanamori.
\end{acknowledgments}

\end{document}